\documentclass[12pt]{article}

\usepackage{amsmath,amssymb,amsfonts,amsthm}
\usepackage{graphicx}
\usepackage{cite}
\usepackage[all]{xy}
\usepackage[toc,page]{appendix}

\allowdisplaybreaks[3]

\newcounter{propositiona}

\newcounter{definitiona}
\newcommand{\definitiona}[1]{\refstepcounter{definitiona}
\noindent
\textbf{Definition \thedefinitiona.}\, #1}
\newcounter{remarka}
\newcommand{\remarka}[1]{\refstepcounter{remarka}
\noindent
\textbf{Remark \theremarka.}\, #1}
\newcounter{examplea}
\newcommand{\examplea}[1]{\refstepcounter{examplea}
\noindent
\textbf{Example \theexamplea.}\, #1}
\newcounter{lemmaa}

\newcounter{theorema}
\newcommand{\theorema}[1]{\refstepcounter{theorema}
\noindent
\textbf{Theorem\, \thetheorema.}\, {\it #1}}
\newcounter{corollarya}

\begin{document}

\thispagestyle{empty}

\begin{center}
{\bf \Large Internal Lagrangians and spatial-gauge symmetries}\\[3.15ex]
{\large \bf Kostya Druzhkov}\\[1.25ex]
Department of Mathematics and Statistics, University of Saskatchewan,\\
Saskatoon, S7N 5E6 Canada\\[0.8ex]

\textit{E-mail: konstantin.druzhkov@gmail.com}
\end{center}

\begin{abstract}\normalsize
A direct reformulation of the Hamiltonian formalism in terms of the intrinsic geometry of infinitely prolonged differential equations is obtained. Concepts of spatial equation and spatial-gauge symmetry of a Lagrangian system of equations are introduced. A non-covariant canonical variational principle is proposed and demonstrated using the Maxwell equations as an example. A covariant canonical variational principle is formulated. 
The results obtained are applicable to any variational equations, including those that do not originate in physics.
\end{abstract}

\noindent
\textit{Keywords: Internal Lagrangians, Presymplectic structures, Hamiltonian formalism, Phase space}

\section{Introduction}{\label{Intro}}

The Lagrangian and Hamiltonian formalisms have their origins in classical mechanics, where the difference between them essentially boils down to the following observation. The Hamiltonian formalism deals with varying an action functional \textit{within the class of all paths through instantaneous states} of a mechanical system. Accordingly, one can interpret it as the Lagrangian formalism rewritten at the level of the intrinsic geometry of equations of motion. The original Lagrangian formalism is formulated in terms of configuration spaces. Another notable difference between these formalisms appears in field theory. Namely, in contrast to the Lagrangian formalism, the Hamiltonian one requires a space+time decomposition to define the meaning of ``instantaneous''. Nonetheless, such decompositions are not based on embeddings of variational systems of equations into jet bundles and can be described in terms of the intrinsic geometry of differential equations. The same applies to classes of all paths through properly defined instantaneous states. In addition, each Lagrangian of a variational system produces a unique internal Lagrangian~\cite{Druzhkov1}, which can be varied within such classes. As a result, some description of variational principles in terms of the intrinsic geometry of infinitely prolonged variational systems arises. This is one of the main ideas of this paper and a simple answer to the following question. \textit{Why does the intrinsic geometry of a variational system know about its variational nature?} By a variational system, we mean a system of differential equations such that some of its non-trivial differential consequences on some finite order jets is the Euler-Lagrange system for a variational problem. Besides, we imply that the corresponding internal Lagrangian must be non-trivial.

Two more questions arise here. \textit{Where does a variational system of equations contain information about its variational nature?} And finally, \textit{how can one interpret geometrical structures that encode such information?} An answer to the second question is given in~\cite{Druzhkov1}. Partial answers to the first and third questions are presented in~\cite{Druzhkov3}. In this paper, we propose more complete answers to the first and third questions. To this end, we describe non-covariant 
phase space formalism in terms of the intrinsic geometry of infinitely prolonged differential equations (for covariant approaches to the Hamiltonian formalism, see, e.g.,~\cite{CrnWit, Got1, Got2, GotIsMa, Helein, Vita, Khavk1, IbSp} and references therein). We introduce notions of spatial equation and spatial-gauge symmetry. Spatial equations are remarkable in that they encode instantaneous states via equivalence classes of their solutions. They deliver a canonical way to introduce instantaneous phase spaces for Lagrangian systems. Internal Lagrangians allow us to formulate a non-covariant canonical variational principle in terms of paths in an instantaneous phase space. The construction applies to all variational equations and does not rely on embedding such equations into jets or the choice of representatives in any equivalence classes. However, it requires fixing the spatial part of a space+time decomposition (the temporal part plays no role). In particular, this construction gives a natural perspective on constrained Hamiltonian systems. In a nutshell, constraints play a \textit{technical} role and arise due to the non-triviality of spatial equations. Finally, the canonical variational principle gives rise to its covariant version.

Broadly speaking, the results obtained are related to the inverse problem of the calculus of variations (see, e.g.,~\cite{Henn, Krup, Khavk2, Grigoriev, Druzhkov2, GriGri}). However, we are focused precisely on
\textbf{how variational principles originating from (action functionals on) jets are encoded in the intrinsic geometry of infinitely prolonged differential equations}. Another way to describe variational principles in terms of the intrinsic geometry of PDEs is proposed in~\cite{Grigoriev} (see also~\cite{GriGri}). The description is based on the concept of \textit{intrinsic} Lagrangian. \textit{Intrinsic} Lagrangians can be considered representatives of \textit{internal} Lagrangians.
This approach allows one not to face constrained variational problems, although variational principles on jets define intrinsic Lagrangians ambiguously.

This paper is organized as follows. In Section~\ref{Basic} we introduce notation and recall some basic concepts from the geometry
of differential equations. Section~\ref{Spaeq} is devoted to spatial equations and spatial-gauge symmetries. Section~\ref{Canovp} recalls a variational principle given by an internal Lagrangian. Section~\ref{Exam} demonstrates the variational principle using several examples. In Section~\ref{Diss} we discuss whether one needs to fix the spatial parts of space+time decompositions to obtain a description of the stationary-action principle in terms of the intrinsic geometry of PDEs. We briefly describe a covariant version of the canonical variational principle.

\textit{All functions and manifolds considered in this paper are assumed to be smooth of the class $C^{\infty}$. All submanifolds are assumed to be embedded.}

\section{Basic notation}{\label{Basic}}

Let us introduce some notation and briefly recall basic facts from the geometry of differential equations. More details can be found in~\cite{VinKr, KraVer1}.

\subsection{Jets}

Let $\pi\colon E\to M$ be a locally trivial smooth vector
bundle over a smooth manifold $M$, $\mathrm{dim}\, M = n$, 
$\mathrm{dim}\, E = n + m$. The bundle $\pi$ gives rise to the corresponding jet bundles $\pi_k\colon J^k(\pi)\to M$,
\begin{align*}
\xymatrix{
\ldots \ar[r] & J^3(\pi) \ar[r]^-{\pi_{3, 2}} & J^2(\pi) \ar[r]^-{\pi_{2, 1}} & J^1(\pi) \ar[r]^-{\pi_{1, 0}} & J^0(\pi) = E \ar[r]^-\pi & M
}
\end{align*}
and the inverse limit $J^{\infty}(\pi)$ arising with the natural projections 
$\pi_{\infty}\colon J^{\infty}(\pi)\to M$ and $\pi_{\infty,\, k}\colon J^{\infty}(\pi)\to J^k(\pi)$. Denote by $\mathcal{F}(\pi)$ the algebra of smooth functions on~$J^{\infty}(\pi)$.\\[-2ex]

\noindent
\textbf{Local coordinates.} Suppose $U\subset M$ is a coordinate neighborhood such that the bundle $\pi$
becomes trivial over $U$. Choose local coordinates $x^1$, \ldots, $x^n$ in $U$ and $u^1$, \ldots, $u^m$  
along the fibers of $\pi$ over $U$. It is convenient to introduce a
multi-index $\alpha$ as a formal sum of the form $\alpha = \alpha_1 x^1 + \ldots + \alpha_n x^n = \alpha_i x^i$, where all $\alpha_i$ are non-negative integers; $|\alpha| = \alpha_1 + \ldots + \alpha_n$.
We denote by $u^i_{\alpha}$ the corresponding adapted local coordinates on $J^{\infty}(\pi)$.\\[-2ex]

\noindent
\textbf{Cartan distribution.} The main structure on jet manifolds is Cartan distribution. If $h$ is a section of $\pi$, there is the section $j^{\infty}h$ of $\pi_{\infty}$ mapping points of $M$ to the respective infinite jets of $h$. The plane $\mathcal{C}_p$ of the Cartan distribution $\mathcal{C}$ at a point $p\in J^{\infty}(\pi)$ is defined as follows. There is a section $h\in \Gamma(\pi)$ such that $p = j^{\infty}h\,(x)$, where $x = \pi_{\infty}(p)$. Then
\begin{align*}
\mathcal{C}_p = d (j^{\infty} h)_x (T_{x} M)\,.
\end{align*}

In local coordinates, the Cartan distribution is spanned by the total derivatives (using the summation convention)
$$
D_{x^i} = \partial_{x^i} + u^k_{\alpha + x^i}\partial_{u^k_{\alpha}}\qquad\quad i = 1, \ldots, n.
$$
We also refer to Cartan planes as horizontal planes. Historically, the concept of Cartan distribution is associated with Johann Friedrich Pfaff, Sophus Lie, etc.\\[-2ex]

\noindent
\textbf{Cartan forms.} The Cartan distribution $\mathcal{C}$ determines the ideal $\mathcal{C}\Lambda^*(\pi)\subset \Lambda^*(\pi)$ 
of the algebra of differential forms on $J^{\infty}(\pi)$.
The ideal $\mathcal{C}\Lambda^*(\pi)$ is generated by Cartan (or contact) forms, i.e., differential forms that annihilate $\mathcal{C}$.
A Cartan $1$-form $\omega\in\mathcal{C}\Lambda^1(\pi)$ can be written as a finite sum
$$
\omega = \omega_i^{\alpha}\theta^i_{\alpha}\,,\qquad\ \theta^i_{\alpha} = du^i_{\alpha} - u^i_{\alpha + x^k}dx^k
$$
in adapted local coordinates. Here $u^i_0 = u^i$, the coefficients $\omega_i^{\alpha}$ are smooth functions defined on a coordinate domain of $J^{\infty}(\pi)$. We denote by $\mathcal{C}^p\Lambda^*(\pi)$ the $p$-th power of the ideal $\mathcal{C}\Lambda^*(\pi)$.\\[-2ex]

\noindent
\textbf{Infinitesimal symmetries.}
The projections $\pi_{\infty,\, k}$ allow one to regard sections of the pullback bundles $\pi_k^*(\pi)$ as sections of the pullback $\pi^*_{\infty}(\pi)$.
Let $\varkappa(\pi) = \Gamma(\pi^*_{\infty}(\pi)) := \bigcup_k \Gamma(\pi^*_{k}(\pi))$ be the $\mathcal{F}(\pi)$-module
of sections of $\pi^*_{\infty}(\pi)$. If $\varphi\in \varkappa(\pi)$, there is the evolutionary vector field on $J^{\infty}(\pi)$
$$
E_{\varphi} = D_{\alpha}(\varphi^i)\partial_{u^i_{\alpha}}\,,
$$
where $\varphi^1$, \ldots, $\varphi^m$ are components of $\varphi$ in adapted local coordinates, $D_{\alpha}$ denotes the composition $D_{x^1}^{\ \alpha_1}\circ\ldots\circ D_{x^n}^{\ \alpha_n}$. Evolutionary vector fields are infinitesimal symmetries of $J^{\infty}(\pi)$. In particular, $\mathcal{L}_{E_{\varphi}}\,\mathcal{C}\Lambda^*(\pi)\subset \mathcal{C}\Lambda^*(\pi)$. Here $\mathcal{L}_{E_{\varphi}}$ is the corresponding Lie derivative.\\[-2ex]

\noindent
\textbf{Horizontal forms.}
Cartan forms allow one to consider the module of horizontal $k$-forms $\Lambda^k_h(\pi) = \Lambda^k(\pi)/\mathcal{C}\Lambda^k(\pi)$. 
The de Rham differential $d$ induces the differential $d_h\colon \Lambda^k_h(\pi)\to \Lambda^{k+1}_h(\pi)$. Each horizontal $n$-form $L$ determines the corresponding 
\textit{horizontal cohomology class} $L + d_h (\Lambda^{n-1}_h(\pi))$.
The infinite jet bundle $\pi_{\infty}\colon J^{\infty}(\pi) \to M$ admits the decomposition 
$$
\Lambda^1(\pi) = \mathcal{C}\Lambda^1(\pi) \oplus \mathcal{F}(\pi)\!\cdot\!\pi^*_{\infty}(\Lambda^1(M))\,.
$$
We identify the module of horizontal $k$-forms $\Lambda^k_h(\pi)$ with $\mathcal{F}(\pi)\cdot \pi^*_{\infty}(\Lambda^k(M))$.\\[-2ex]

\noindent
\textbf{Euler operator.}
By $\mathrm{E}$ we denote the Euler operator (variational derivative), $\mathrm{E}\colon \Lambda^n_h(\pi)\to \mathrm{Hom}_{\mathcal{F}(\pi)}(\varkappa(\pi), \Lambda^n_h(\pi))$. 
In local coordinates, for $L = \lambda \, dx^1\wedge\ldots\wedge dx^n$, we have
\begin{align*}
&\langle \mathrm{E}(L), \varphi \rangle = \dfrac{\delta \lambda}{\delta u^i}\,\varphi^i\wedge dx^1\wedge\ldots\wedge dx^n,\qquad \dfrac{\delta \lambda}{\delta u^i} = (-1)^{|\alpha|}D_{\alpha}\Big(\dfrac{\partial \lambda}{\partial u^i_{\alpha}}\Big).
\end{align*}
We also regard $\mathrm{E}(L)$ as the differential form
\begin{align*}
&\mathrm{E}(L) = \dfrac{\delta \lambda}{\delta u^i}\,\theta_0^i\wedge dx^1\wedge\ldots\wedge dx^n.
\end{align*}

\subsection{Differential equations \label{difeq}}

Let $\eta$ be a locally trivial smooth vector bundle over the same base as $\pi$, and let $F$ be a (smooth) section of a bundle of the form $\pi^*_{r}(\eta)$. We assume that, for each point $p\in \{F = 0\}\subset J^{r}(\pi)$, the differentials $dF^i_p$ of the coordinate functions are linearly independent. By infinite prolongation of the differential equation $F = 0$, we mean the subset $\mathcal{E}\subset J^{\infty}(\pi)$ that is defined by the infinite system of equations
\begin{align*}
\quad D_{\alpha}(F^i) = 0\,,\qquad |\alpha| \geqslant 0\,.
\end{align*}
Henceforth, we denote by $\pi_{\mathcal{E}}$ the restriction of $\pi_{\infty}$ to $\mathcal{E}$ and assume that $\pi_{\mathcal{E}}(\mathcal{E}) = M$.

\vspace{1.5ex}

\remarka{We do not require that the number of equations of the form $F^i = 0$ coincide with the number of dependent variables.}

\vspace{1.5ex}

By $\mathcal{F}(\mathcal{E})$ we denote the algebra of smooth functions on $\mathcal{E}$, 
$$
\mathcal{F}(\mathcal{E}) = \mathcal{F}(\pi)|_{\mathcal{E}} = \mathcal{F}(\pi)/I.
$$
Here $I$ denotes the ideal of the system $\mathcal{E}\subset J^{\infty}(\pi)$.\\[-2ex]

\noindent
\textbf{Regularity assumptions.}
We say that the infinite prolongation $\mathcal{E}$ of a system of differential equations $F = 0$ is \textit{regular} if for every
function $f\in \mathcal{F}(\pi)$ that vanishes on $\mathcal{E}$ (i.e., $f\in I$), there exists an operator in total derivatives $\Delta\colon P(\pi)\to \mathcal{F}(\pi)$ such that $f = \Delta(F)$.
\textit{In what follows, we
consider only regular systems.}\\[-2ex]

The algebra of smooth functions produces the algebra of differential forms $\Lambda^*(\mathcal{E}) = \Lambda^*(\pi)|_{\mathcal{E}}$. The Cartan distribution on $J^{\infty}(\pi)$ can be restricted to $\mathcal{E}$.
Similarly, there is the ideal $\mathcal{C}\Lambda^*(\mathcal{E})\subset \Lambda^*(\mathcal{E})$ generated by differential forms that annihilate the Cartan distribution on $\mathcal{E}$. Note that $\mathcal{C}\Lambda^*(\mathcal{E}) = \mathcal{C}\Lambda^*(\pi)|_{\mathcal{E}}$ due to the decomposition of $\Lambda^1(\pi)$. A \textit{solution} of $\pi_{\mathcal{E}}$ is a section $\sigma\colon M \to \mathcal{E}$ such that for each $x\in M$, 
$$
d\sigma_x(T_x M) = \mathcal{C}_{\sigma(x)}\,.
$$

Let $P(\pi)$ be the module of sections of the pullback $\pi_{\infty}^*(\eta)$. Introduce the $\mathcal{F}(\mathcal{E})$-modules
$$
\varkappa(\mathcal{E}) = \varkappa(\pi)/I\cdot \varkappa(\pi)\,,\qquad\quad P(\mathcal{E}) = P(\pi)/I\cdot P(\pi)\,.
$$

\noindent
\textbf{Infinitesimal symmetries.} A \textit{symmetry} (more precisely, infinitesimal symmetry) of an infinitely prolonged system of equations $\pi_{\mathcal{E}}$ is a $\pi_{\mathcal{E}}$-vertical vector field $X\in D(\mathcal{E})$ that preserves the Cartan distribution, i.e., $\mathcal{L}_X\, \mathcal{C}\Lambda^*(\mathcal{E})\subset \mathcal{C}\Lambda^*(\mathcal{E})$.

A differential equation $\{F = 0\}\subset J^{r}(\pi)$ allows one to introduce the linearization $l_{\mathcal{E}}\colon \varkappa(\mathcal{E})\to P(\mathcal{E})$. Namely, let us define the operator $l_F\colon \varkappa(\pi)\to P(\pi)$ by $l_F(\varphi) = E_{\varphi}(F)$ and set $l_{\mathcal{E}} = l_F|_{\mathcal{E}}$.
If $\pi_{\infty,\, 0}(\mathcal{E}) = J^0(\pi)$, symmetries of $\pi_{\mathcal{E}}$ can be identified with elements of $\ker l_{\mathcal{E}}$ using their characteristics (see, e.g.,~\cite{VinKr},~\cite{Olver}).
A \textit{gauge symmetry} of $\pi_{\mathcal{E}}$ is a symmetry of the form $R(\epsilon)\in \ker l_{\mathcal{E}}$, where $R$ is an operator in total derivatives such that $l_{\mathcal{E}}\circ R = 0$.\\[-2ex]

\noindent
\textbf{$\mathcal{C}$-spectral sequence.} Powers of the ideal $\mathcal{C}\Lambda^*(\mathcal{E})$ are stable with respect to the de Rham differential, where $\mathcal{C}^{p+1}\Lambda^p(\mathcal{E}) = 0$.
Then, the de Rham complex admits the filtration
$$
\Lambda^{\bullet}(\mathcal{E})\supset \mathcal{C}\Lambda^{\bullet}(\mathcal{E})\supset \mathcal{C}^2\Lambda^{\bullet}(\mathcal{E})\supset \mathcal{C}^3\Lambda^{\bullet}(\mathcal{E})\supset \ldots
$$
The corresponding spectral sequence $(E^{p,\, q}_r(\mathcal{E}), d^{\,p,\, q}_r)$ is the Vinogradov $\mathcal{C}$-spectral sequence~\cite{Vin, VinKr}.
Here $d^{\,p,\, q}_r\colon E^{p,\, q}_r(\mathcal{E})\to E^{p+r,\, q+1-r}_r(\mathcal{E})$ are induced by the de Rham differential $d$,
\begin{align*}
&E^{p,\, q}_0(\mathcal{E}) = \mathcal{C}^p\Lambda^{p+q}(\mathcal{E})/\mathcal{C}^{p+1}\Lambda^{p+q}(\mathcal{E})\,,\qquad 
E^{p,\, q}_1(\mathcal{E}) = \ker d_0^{\,p,\, q}/\mathrm{im}\, d_0^{\,p,\, q-1}\,,\qquad \ldots
\end{align*}
In particular, $\mathcal{C}$-spectral sequence allows one to define conservation laws, variational $1$-forms and presymplectic structures of differential equations.
A \textit{variational $1$-form} of $\mathcal{E}$ is an element of the group $E^{\,1,\,n-1}_1(\mathcal{E})$.
A \textit{presymplectic structure} of $\mathcal{E}$ is an element of the kernel of the differential
$$
d_1^{\,2,\,n-1}\colon E^{\,2,\,n-1}_1(\mathcal{E})\to E^{\,3,\,n-1}_1(\mathcal{E}).
$$
We identify each group $E^{p,\, q}_1(\mathcal{E})$ with its canonically isomorphic group
$$
\dfrac{\{\omega\in \mathcal{C}^p\Lambda^{p+q}(\mathcal{E})\colon d\omega \in \mathcal{C}^{p+1}\Lambda^{p+q+1}(\mathcal{E})\}}{\mathcal{C}^{p+1}\Lambda^{p+q}(\mathcal{E}) + d(\mathcal{C}^{p}\Lambda^{p+q-1}(\mathcal{E}))}\,.
$$

\noindent
\textbf{Internal Lagrangians (see~\cite{Druzhkov1, Druzhkov3}).} 
Suppose that the variational derivative $\mathrm{E}(L)$ of a 
horizontal $n$-form $L\in\mathcal{F}(\pi)\cdot \pi_{\infty}^*(\Lambda^n(M))$ vanishes on $\mathcal{E}$.
There exists a form $\omega_L\in\mathcal{C}\Lambda^{n}(\pi)$ such that $d(L + \omega_L) - \mathrm{E}(L) \in\mathcal{C}^2\Lambda^{n+1}(\pi)$. Although such a form $\omega_L$ is defined ambiguously, all differential forms of the form $(L + \omega_L)|_{\mathcal{E}}$ represent the same element of the group
\begin{align}
\dfrac{\{l\in \Lambda^n(\mathcal{E})\, \colon \ dl \in \mathcal{C}^2\Lambda^{n+1}(\mathcal{E})\}}{\mathcal{C}^2\Lambda^{n}(\mathcal{E}) + d(\mathcal{C}\Lambda^{n-1}(\mathcal{E}))}\,.
\label{IntSepLag}
\end{align}
Thus, $L$ defines a unique element of group~\eqref{IntSepLag}. For example, one can take any presymplectic potential current as $\omega_L$, i.e., any $\omega_L = {\omega_L}^{\alpha k}_{\, i}\hspace{0.1ex}\theta^i_{\alpha}\wedge (\partial_{x^k}\, \lrcorner\, dx^1\wedge \ldots\wedge dx^n)$ such that for each $\varphi\in \varkappa(\pi)$, 
$$
\mathcal{L}_{E_{\varphi}} L = \langle \mathrm{E}(L), \varphi \rangle + d_h (E_{\varphi}\, \lrcorner \, \omega_L)\,.
$$
It can be derived using integration by parts.

Similarly, the horizontal cohomology class of $L$ defines a unique \textit{internal Lagrangian}, i.e., element of the group
\begin{align*}
\widetilde{E}^{\,0,\, n-1}_1(\mathcal{E}) = \dfrac{\{l\in\Lambda^n(\mathcal{E})\colon\ dl\in \mathcal{C}^2\Lambda^{n+1}(\mathcal{E})\}}
{\mathcal{C}^2\Lambda^{n}(\mathcal{E}) + d(\Lambda^{n-1}(\mathcal{E}))}\,.
\end{align*}
This group appears in the spectral sequence for Lagrangian formalism~\cite{Druzhkov1}, which is produced by the filtration
$
\Lambda^{\bullet}(\mathcal{E})\supset \mathcal{C}^2\Lambda^{\bullet}(\mathcal{E})\supset \mathcal{C}^3\Lambda^{\bullet}(\mathcal{E}) \supset \mathcal{C}^4\Lambda^{\bullet}(\mathcal{E})\supset \ldots
$
of the de Rham complex (hence the notation). The de Rham differential $d$ induces the differential
$$
\tilde{d}^{\,0,\,n-1}_1\colon \widetilde{E}^{\,0,\, n-1}_1(\mathcal{E})\to E^{\,2,\, n-1}_1(\mathcal{E})\,,
$$
which maps internal Lagrangians to presymplectic structures (i.e., $\mathrm{im}\, \tilde{d}^{\,0,\,n-1}_1\subset \ker d_1^{\,2,\,n-1}$).

Let us shed some light on the indexing, as it is rather specific. Internal Lagrangians are elements of the cohomology of
\begin{align*}
\dfrac{\Lambda^{n-1}(\mathcal{E})}{\mathcal{C}^2\Lambda^{n-1}(\mathcal{E})}\to \dfrac{\Lambda^{n}(\mathcal{E})}{\mathcal{C}^2\Lambda^{n}(\mathcal{E})}\to \dfrac{\Lambda^{n+1}(\mathcal{E})}{\mathcal{C}^2\Lambda^{n+1}(\mathcal{E})}\,.
\end{align*}
Elements of the quotient $\Lambda^{n}(\mathcal{E})/\mathcal{C}^2\Lambda^{n}(\mathcal{E})$ can be unambiguously restricted to tangent planes that are spanned by $n-1$ horizontal vectors and $1$ vector of unspecified (any) type.

If an infinitely prolonged system $\mathcal{E}$ is embedded into some jets $J^{\infty}(\pi)$, then each element of group~\eqref{IntSepLag} ambiguously defines a horizontal $n$-form $L$ such that $\mathrm{E}(L)|_{\mathcal{E}} = 0$ (see~\cite{Druzhkov1}, Theorem 1). In essence, this fact is based on the results obtained in~\cite{Khavk2} (Theorem 3).

\section{Spatial equations}{\label{Spaeq}}

Let $\pi_{\mathcal{E}}\colon \mathcal{E}\to M^n$ be an infinitely prolonged system of differential equations. The Cartan distribution $\mathcal{C}$ of $\mathcal{E}$ can be considered a connection (horizontal distribution). By a \textit{spatial distribution} on $\mathcal{E}$, we mean the lift of an involutive hyperplane distribution from $M$ to $\mathcal{E}$. We prefer to focus on regular distributions, but one can also consider some singular ones.

\vspace{1.5ex}

\remarka{There is the tangent distribution on the boundary $\partial M$ and its horizontal lift (over the boundary) known as the boundary equation. In this paper, we deal with the lifts of distributions from the entire base $M$, but one can say that there are similar ideas behind these constructions.}

\vspace{1.5ex}

Suppose $\mathcal{S}$ is a spatial distribution. One can informally say that $\mathcal{S}$ endows $\mathcal{E}$ with another structure of a differential equation since it is also involutive. We call such equations spatial (or instantaneous states) equations. An $(n - 1)$-dimensional (embedded) integral manifold of $\mathcal{S}$ determines a \textit{solution to the spatial equation} if its projection to $M$ is embedding. Solutions to a spatial equation encode (local) instantaneous states.

A spatial equation $(\mathcal{E}, \mathcal{S})$ produces the corresponding Vinogradov's spectral sequence, which we call the $\mathcal{S}$-spectral one. 
Let $\mathcal{S}\Lambda^*(\mathcal{E}) \subset \Lambda^*(\mathcal{E})$ be the ideal generated by differential forms that annihilate $\mathcal{S}$. Denote by $\mathcal{S}^k\Lambda^*(\mathcal{E})$ the $k$-th power of $\mathcal{S}\Lambda^*(\mathcal{E})$. Since planes of the Cartan distribution $\mathcal{C}$ contain the respective planes of $\mathcal{S}$, one obtains the dual inclusion $\mathcal{S}\Lambda^1(\mathcal{E})\supset \mathcal{C}\Lambda^1(\mathcal{E})$, which also implies that
\begin{align*}
\mathcal{S}^p\Lambda^{p+q}(\mathcal{E})\supset \mathcal{C}^p\Lambda^{p+q}(\mathcal{E})\,.
\end{align*}
The inclusions $\mathcal{S}^p\Lambda^{\bullet}(\mathcal{E})\supset \mathcal{C}^p\Lambda^{\bullet}(\mathcal{E})$ of the complexes give rise to the corresponding morphism between $\mathcal{C}$-spectral sequence and $\mathcal{S}$-spectral one. To name $\mathcal{S}$-spectral sequence terms, we use the prefix ``$\mathcal{S}$-'' and the names of $\mathcal{C}$-spectral sequence terms that have the same structure of indices. For example, each variational $1$-form $\omega + \mathcal{C}^2\Lambda^{n}(\mathcal{E}) + d(\mathcal{C}\Lambda^{n-1}(\mathcal{E}))$ determines the corresponding $\mathcal{S}$-variational $1$-form 
$$
\omega + \mathcal{S}^2\Lambda^{n}(\mathcal{E}) + d(\mathcal{S}\Lambda^{n-1}(\mathcal{E}))\,.
$$
Let us note that $\mathcal{S}$-variational $1$-forms are not variational $1$-forms for the spatial equation $(\mathcal{E}, \mathcal{S})$ since its variational $1$-forms are represented by differential $(n-1)$-forms.

Similarly, we call elements of the group 
\begin{align*}
\dfrac{\{l\in \Lambda^n(\mathcal{E})\,\colon\ dl\in \mathcal{S}^2\Lambda^{n+1}(\mathcal{E})\}}{\mathcal{S}^2\Lambda^n(\mathcal{E}) + d(\Lambda^{n-1}(\mathcal{E}))}
\end{align*}
$\mathcal{S}$-internal Lagrangians. An internal Lagrangian $\boldsymbol \ell = l + \mathcal{C}^2\Lambda^n(\mathcal{E}) + d(\Lambda^{n-1}(\mathcal{E}))$ defines the corresponding $\mathcal{S}$-internal Lagrangian $l + \mathcal{S}^2\Lambda^n(\mathcal{E}) + d(\Lambda^{n-1}(\mathcal{E}))$.
An $\mathcal{S}$-internal Lagrangian $l + \mathcal{S}^2\Lambda^n(\mathcal{E}) + d(\Lambda^{n-1}(\mathcal{E}))$ determines the corresponding $\mathcal{S}$-presymplectic structure $dl + \mathcal{S}^3\Lambda^{n+1}(\mathcal{E}) + d(\mathcal{S}^2\Lambda^{n}(\mathcal{E}))$.

\vspace{1.5ex}

\remarka{Since $\Lambda^n(\mathcal{E}) = \mathcal{S}\Lambda^n(\mathcal{E})$, each element of group~\eqref{IntSepLag} determines a unique $\mathcal{S}$-variational $1$-form. This agrees with the Lagrangian formalism in classical mechanics, where Lagrangians are differential $1$-forms. Within this analogy, $\mathcal{S}$-internal Lagrangians play the role of horizontal cohomology classes of Lagrangians.}

\vspace{1.5ex}

\definitiona{Let $\mathcal{S}$ be a spatial distribution on $\mathcal{E}$. A $\pi_{\mathcal{E}}$-vertical vector field $X$ on $\mathcal{E}$ is an $\mathcal{S}$\textit{-symmetry} if 
$$
\mathcal{L}_X\, \mathcal{S}\Lambda^*(\mathcal{E})\subset \mathcal{S}\Lambda^*(\mathcal{E})\,.
$$}
Symmetries of $\mathcal{E}$ are indifferent to the choice of a spatial distribution, i.e., if a $\pi_{\mathcal{E}}$-vertical vector field is a symmetry of $\mathcal{E}$, it is an $\mathcal{S}$-symmetry for any spatial distribution $\mathcal{S}$.

In some cases, it is advisable to regard spatial equations as gauge systems. For such spatial equations, (local) instantaneous states should be deemed not just their solutions but equivalence classes of solutions. Then, a notion of spatial-gauge symmetry is required. Perhaps $\mathcal{S}$-gauge symmetries can be defined in various reasonable ways. For Euler-Lagrange equations, we propose the concept of $(\boldsymbol \ell, \mathcal{S})$-gauge symmetry.

\vspace{1.5ex}

\definitiona{Let $\mathcal{S}$ be a spatial distribution on a system of differential equations $\pi_{\mathcal{E}}$, and let $\boldsymbol \ell$ be an internal Lagrangian of $\mathcal{E}$. An $\mathcal{S}$-symmetry is an $(\boldsymbol \ell, \mathcal{S})$\textit{-gauge symmetry} if substituting it into the corresponding $\mathcal{S}$-presymplectic structure yields the trivial $\mathcal{S}$-variational $1$-form.}

\vspace{1.5ex}

\noindent
This way of defining $\mathcal{S}$-gauge symmetries is analogous to how one can equivalently define gauge symmetries for Lagrangian systems. 

If an $(\boldsymbol \ell, \mathcal{S})$-gauge symmetry generates a global flow\footnotemark[1], the corresponding transformations play the role of spatial-gauge ones. For a spatial distribution $\mathcal{S}$, the set of all such transformations generates a group (with composition as the group operation), which we call $(\boldsymbol \ell, \mathcal{S})$\textit{-gauge group}.

\footnotetext[1]{An $\mathcal{S}$-symmetry can define transformations of solutions to $\mathcal{S}$ (and of $\mathcal{S}$-sections, see Section \ref{Canovp}) by means of the flow of an equivalent non-vertical symmetry of $\mathcal{S}$, i.e., vector field on $\mathcal{E}$ such that at each point $x\in \mathcal{E}$, it differs from the $\mathcal{S}$-symmetry by a vector lying in the plane $\mathcal{S}_x$.}

Let us recall that substituting a gauge symmetry of $\mathcal{E}$ into a presymplectic structure yields the trivial variational $1$-form corresponding to the trivial $\mathcal{S}$-variational $1$-form. Hence, we obtain

\vspace{1.5ex}

\theorema{Let $\mathcal{S}$ be a spatial distribution on a system of differential equations $\pi_{\mathcal{E}}$, and let $\boldsymbol \ell$ be an internal Lagrangian of $\mathcal{E}$. Then any gauge symmetry of $\pi_{\mathcal{E}}$ is an $(\boldsymbol \ell, \mathcal{S})$-gauge symmetry.}

\vspace{1.5ex}

The next theorem also follows immediately from the definitions.

\vspace{1.5ex}

\theorema{Let $\mathcal{S}$ be a spatial distribution on a system of equations $\pi_{\mathcal{E}}$, and let $\boldsymbol \ell$ be an internal Lagrangian of $\mathcal{E}$. Then the corresponding $\mathcal{S}$-internal Lagrangian is $(\boldsymbol \ell, \mathcal{S})$-gauge invariant.}\\[0.5ex]
\textbf{Proof.} Suppose $X$ is an $(\boldsymbol \ell, \mathcal{S})$-gauge symmetry. If $l\in \Lambda^n(\mathcal{E})$ represents $\boldsymbol \ell$, then the $\mathcal{S}$-internal Lagrangian $\mathcal{L}_X l + \mathcal{S}^2\Lambda^{n}(\mathcal{E}) + d(\Lambda^{n-1}(\mathcal{E}))$ is represented by the differential form $X\lrcorner\, dl$. Since $dl$ produces the desired $\mathcal{S}$-presymplectic structure, $X\lrcorner\, dl \in \mathcal{S}^2\Lambda^{n}(\mathcal{E}) + d(\mathcal{S}\Lambda^{n-1}(\mathcal{E}))\subset \mathcal{S}^2\Lambda^{n}(\mathcal{E}) + d(\Lambda^{n-1}(\mathcal{E}))$, and hence, the $\mathcal{S}$-internal Lagrangian $\mathcal{L}_X l + \mathcal{S}^2\Lambda^{n}(\mathcal{E}) + d(\Lambda^{n-1}(\mathcal{E}))$ is trivial. 

\vspace{1.5ex}

We use more suitable indices for local coordinates on jets and equations in the examples below.

\vspace{1.5ex}

\examplea{Let us consider the infinite prolongation $\mathcal{E}$ of the Laplace equation
\begin{align*}
u_{yy} = -u_{xx}
\end{align*}
and its internal Lagrangian represented by the differential form $l = (L + \omega_L)|_{\mathcal{E}}$ (e.g. \cite{Druzhkov3}, Example~2),
\begin{align*}
L + \omega_L = -\frac{u_x^2 + u_y^2}{2}\,dx\wedge dy - u_x\,\theta_0\wedge dy + u_y\,\theta_0\wedge dx\,.
\end{align*}
Here $\theta_0$ denotes $du - u_x dx - u_y dy$; $\pi\colon \mathbb{R}\times \mathbb{R}^2\to \mathbb{R}^2$ is the projection onto the second factor.

One can regard the coordinate $u_{yy}$ and its derivatives as external coordinates for the infinite prolongation. Other coordinates on $J^{\infty}(\pi)$ can be treated as local coordinates on $\mathcal{E}$. Then the restrictions of the total derivatives to the system $\mathcal{E}$ have the form
\begin{align*}
&\,\overline{\!D}_x = \partial_x + u_x\partial_u + u_{xx}\partial_{u_x} + u_{xy}\partial_{u_y} + u_{xxx}\partial_{u_{xx}} + u_{xxy}\partial_{u_{xy}} + u_{xxxx}\partial_{u_{xxx}} + \ldots\,,\\
&\,\overline{\!D}_y = \partial_y + u_y\partial_u + u_{xy}\partial_{u_x} - u_{xx}\partial_{u_y} + u_{xxy}\partial_{u_{xx}} - u_{xxx}\partial_{u_{xy}} + u_{xxxy}\partial_{u_{xxx}} + \ldots
\end{align*}

Suppose $\mathcal{S}$ is the lift of the distribution $\ker dy$. Then at each point of $\mathcal{E}$, the spatial distribution $\mathcal{S}$ is spanned by the total derivative $\,\overline{\!D}_x$. Solutions to the spatial equation have the form
\begin{align*}
&y = y_0\,,\qquad u = f(x)\,,\qquad u_y = g(x)\,,\qquad u_x = \partial_x f\,,\qquad u_{xy} = \partial_x g\,,\qquad \ldots
\end{align*}
Here $y_0\in \mathbb{R}$, $f, g$ are arbitrary functions of $x$.

The presymplectic structure is represented by the form $dl$,
\begin{align*}
dl = - \,\overline{\!\theta}_x\wedge \,\overline{\!\theta}_0\wedge dy + \,\overline{\!\theta}_y\wedge\,\overline{\!\theta}_0\wedge dx\,.
\end{align*}
Here $\,\overline{\!\theta}_0 = \theta_0|_{\mathcal{E}}$,\  $\,\overline{\!\theta}_x = du_x - u_{xx} dx - u_{xy} dy$,\ $\,\overline{\!\theta}_y = du_y - u_{xy} dx + u_{xx} dy$. One can see that the differential form
\begin{align*}
\omega = \,\overline{\!\theta}_y\wedge\,\overline{\!\theta}_0\wedge dx
\end{align*}
represents the same $\mathcal{S}$-presymplectic structure as $dl$ due to the membership $ \,\overline{\!\theta}_x\wedge \,\overline{\!\theta}_0\wedge dy \in \mathcal{S}^3\Lambda^{3}(\mathcal{E})$.

Any $\mathcal{S}$-symmetry has the form
\begin{align*}
X = \varphi\partial_u + \chi\partial_{u_y} + \,\overline{\!D}_x(\varphi)\partial_{u_x} + \,\overline{\!D}_x(\chi)\partial_{u_{xy}} + \ldots
\end{align*}
Here $\varphi$ and $\chi$ are arbitrary functions on $\mathcal{E}$. Then
\begin{align*}
X \lrcorner\, \omega = \chi \ \overline{\!\theta}_0\wedge dx - \varphi\ \overline{\!\theta}_y\wedge dx\,.
\end{align*}
Denote the coordinate $u_y$ by $v$. The spatial equation is isomorphic to the infinite prolongation of the underdetermined ODE system given by just one equation $y_x = 0$ for three dependent variables $y, u, v$. Hence, one can use the results of~\cite{Verb} (Corollary $3.3.$) to show that a differential form 
$$
a\, dy\wedge dx + b \ \overline{\!\theta}_0\wedge dx + c \ \overline{\!\theta}_y\wedge dx
$$ 
represents the trivial $\mathcal{S}$-variational $1$-form if and only if $b = c = 0$ and there is a function $\epsilon\in \mathcal{F}(\mathcal{E})$ such that $a = \,\overline{\!D}_x(\epsilon)$.

So, $X \lrcorner\, \omega$ represents the trivial $\mathcal{S}$-variational $1$-form if and only if $\varphi = \chi = 0$. In the case under consideration, there are no non-trivial $(\boldsymbol \ell, \mathcal{S})$-gauge symmetries.
}

\vspace{1.5ex}

\examplea{\label{MaxExample}
Let $\pi\colon \mathbb{R}^n\times \mathbb{R}^n \to \mathbb{R}^n$ be the projection onto the second factor with coordinates $t = x^0, x^1, \ldots, x^{n-1}$ on the base and with $A^0, \ldots, A^{n-1}$ as coordinates along the fibers, $n > 2$.

The Maxwell equations in vacuum have the form
\begin{align}
\partial_{\mu} F^{\mu\nu} = 0\,.
\label{Max}
\end{align}
Here $F^{\mu\nu}$ denotes $\partial^{\mu}A^{\nu} - \partial^{\nu}A^{\mu}$; the metric $\mathrm{diag}(+1, -1, \ldots, -1)$ is used to rise and lover indices. We assume that the indices $\mu$ and $\nu$ can take all the values $0$, \ldots, $n - 1$. It is convenient to use $i, j, k$ as spatial indices ($i, j, k = 1, \ldots, n-1$).

Denote by $\mathcal{E}$ the infinite prolongation of system~\eqref{Max}. The Lagrangian
\begin{align*}
L = - \dfrac{1}{4} F_{\mu\nu}F^{\mu\nu} d^nx\,,\qquad d^nx = dx^0\wedge \ldots \wedge dx^{n-1}
\end{align*}
gives rise to the internal Lagrangian $\boldsymbol \ell$ represented by the differential form $l = (L + \omega_L)|_{\mathcal{E}}$, where
\begin{align*}
&L + \omega_L = - \dfrac{1}{4} F_{\mu\nu}F^{\mu\nu} d^nx - F_{\mu \nu}\theta^{\nu}\wedge (\partial^{\mu} \lrcorner\, d^nx)\,,\qquad \theta^{\nu} = dA^{\nu} - \partial_{\mu}A^{\nu} dx^{\mu}\,.
\end{align*}

For $\mathcal{S}$, we choose the lift of the distribution $\ker dt$ from the base to $\mathcal{E}$.
The form $dl$ represents the same $\mathcal{S}$-presymplectic structure as
\begin{align}
&\omega = -\,\overline{\!\theta}_{0i}\wedge \,\overline{\!\theta}^{i}\wedge ({\partial^0} \lrcorner\, d^nx),
\label{Maxpres}
\end{align}
where $\,\overline{\!\theta}_{0i} = (dF_{0i} - \partial_{\mu}F_{0i} dx^{\mu})|_{\mathcal{E}}$, and $\,\overline{\!\theta}^i = \theta^i|_{\mathcal{E}}$.

Suppose now $F^{0i}$ is not just a notation but additional dependent variables. Then, using the corresponding jets, $\mathcal{E}$ can be rewritten as the infinite prolongation of the system
\begin{align*}
\begin{aligned}
&F^{0i} = \partial^0 A^i - \partial^i A^0,\\
&\partial_0 F^{0i} = \partial_j(\partial^{i}A^{j} - \partial^{j}A^{i})\,,\\
&\partial_i F^{0i} = 0\,.
\end{aligned}
\end{align*}
More precisely, this infinite prolongation defines the embedding of $\mathcal{E}$ into the corresponding jets, but we identify it with $\mathcal{E}$. So, in addition to $x^{\mu}$, we can treat $A^{\nu}$, $F^{0i}$, $\partial_0 A^0$, $\partial_0^2 A^0, \ldots$ and their spatial derivatives as coordinates on $\mathcal{E}$, except for, say, $\partial_{1}F^{01}$ and its spatial derivatives. One can now see that any $\mathcal{S}$-symmetry of $\mathcal{E}$ has the form
\begin{align}
X_{(\chi, \eta, \varphi)} = \chi^i\partial_{A^i} + \eta^i\partial_{F^{0i}} + \varphi^0\partial_{A^0} + \varphi^1\partial_{\partial_0 A^0} + \varphi^2\partial_{\partial^2_0 A^0} + \ldots
\label{MaxSsym}
\end{align}
Here $\chi^i, \varphi^0, \varphi^1, \ldots \in \mathcal{F}(\mathcal{E})$ can be chosen arbitrarily, while $\eta^i \in \mathcal{F}(\mathcal{E})$ satisfy the relation $\,\overline{\!D}_i(\eta^i) = 0$; $\,\overline{\!D}_{k} = D_k|_{\mathcal{E}}$. Such functions unambiguously define the corresponding $\mathcal{S}$-symmetry. The action of the $X_{(\chi, \eta, \varphi)}$ on derivatives is defined uniquely due to the requirement of commuting with the spatial total derivatives. We obtain
\begin{align*}
X_{(\chi, \eta, \varphi)}\, \lrcorner\, \omega = - \eta^i \, \overline{\!\theta}_{i}\wedge ({\partial^0} \lrcorner\, d^nx) + \chi^i \,\overline{\!\theta}_{0i}\wedge ({\partial^0} \lrcorner\, d^nx)\,.
\end{align*}
By applying the results of~\cite{Verb} (Corollary $3.3.$) to the spatial equation, one can show that this differential form represents the trivial $\mathcal{S}$-variational $1$-form if and only if $\eta^i = 0$ and there exists a function $\epsilon \in \mathcal{F}(\mathcal{E})$ such that
\begin{align*}
\chi^i = \,\overline{\!D}^{\,i}(\epsilon)\,.
\end{align*}
Thus, $(\boldsymbol \ell, \mathcal{S})$-gauge symmetries of Maxwell's equations have the form~\eqref{MaxSsym} for $\chi^i = \,\overline{\!D}^{\,i}(\epsilon)$, $\eta^i = 0$.

}

\section{Non-covariant canonical variational principle}{\label{Canovp}}

Let us recall how internal Lagrangians encode variational principles~\cite{Druzhkov3}.

\vspace{1.5ex}

Consider a system of differential equations $\pi_{\mathcal{E}}\colon \mathcal{E}\to M^n$ and an involutive hyperplane distribution $s$ on its base $M^n$. Let $\boldsymbol \ell$ be an internal Lagrangian of $\mathcal{E}$ represented by a differential form $l\in\Lambda^n(\mathcal{E})$. The lift $\mathcal{S}$ of the distribution $s$ from $M$ allows one to introduce the following concepts.

\vspace{1.5ex}

\definitiona{A section $\sigma$ of the bundle $\pi_{\mathcal{E}}$ is an $\mathcal{S}$\textit{-section} if for each $x\in M$, 
\begin{align*}
d\sigma_x (s_{x}) = \mathcal{S}_{\sigma(x)}.
\end{align*}
}
One can say that $\mathcal{S}$-sections encode paths through instantaneous states determined by the corresponding spatial equation. Generally speaking, an integral manifold of $s$ is not necessarily embedded submanifold of $M$ (because of its possible topology). In a neighborhood of an interior point of $M$, one can ambiguously introduce a parameter (an analog of time) and treat integral manifolds of $s$ as Cauchy surfaces. The concept of $\mathcal{S}$-sections allows us to formulate a global version of the variational principle.

\vspace{1.5ex}

\definitiona{A smooth mapping $\gamma\colon \mathbb{R}\times M\to \mathcal{E}$ is a \textit{path in} $\mathcal{S}$\textit{-sections} if the mappings
\begin{align*}
\gamma(\tau)\colon x\mapsto \gamma(\tau, x)
\end{align*}
are $\mathcal{S}$-sections for all $\tau\in\mathbb{R}$.
}

\vspace{1.5ex}

\definitiona{An $\mathcal{S}$-section $\sigma$ is an $\mathcal{S}$\textit{-stationary point} of $\boldsymbol \ell$ (or a stationary point of the corresponding $\mathcal{S}$-internal Lagrangian) if for any compact oriented $n$-dimensional submanifold $N\subset M$, the relation
\begin{align*}
\dfrac{d}{d\tau}\Big|_{\tau = 0}\int_N \gamma(\tau)^*(l) = 0
\end{align*}
holds for each path $\gamma$ in $\mathcal{S}$-sections such that $\gamma(0) = \sigma$ and all points of the boundary $\partial N$ are fixed (i.e., for each $x\in\partial N$, the condition $\gamma(\tau, x) = \gamma(0, x)$ is satisfied for all $\tau\in\mathbb{R}$).
}

\vspace{1.5ex}

\remarka{If $\boldsymbol \Omega$ is an $\mathcal{S}$-variational $1$-form and for each $x\in \partial M$, $s_x = T_x\, \partial M$, then the action
\begin{align*}
\sigma\mapsto \int_M \sigma^*(\boldsymbol \Omega)
\end{align*}
is well-defined on $\mathcal{S}$-sections, provided $M$ is compact and oriented.
}

\vspace{1.5ex}

All solutions of a system of differential equations are $\mathcal{S}$-stationary points for any of its internal Lagrangians and any spatial distribution taken as $\mathcal{S}$.
Let us also stress that \textit{the concept of $\mathcal{S}$-stationary points of an internal Lagrangian does not depend on the choice of a representative}. This concept leads to desired results when $\mathcal{S}$ is the lift of a nowhere characteristic distribution. More specifically, the following theorem holds~\cite{Druzhkov3} (Theorem 2 in slightly different terminology).

\vspace{1.5ex}

\theorema{Let $L\in \Lambda^n(J^r(\pi))$ be a horizontal differential form, and let $\mathcal{E}$ be the infinite prolongation of the corresponding system of Euler-Lagrange equations. Suppose $\mathcal{S}$ is the lift of a nowhere characteristic involutive hyperplane distribution. Then an $\mathcal{S}$-section $\sigma$ is an $\mathcal{S}$-stationary point of the corresponding internal Lagrangian if and only if $\sigma$ is a solution to $\pi_{\mathcal{E}}$.}

\vspace{1.5ex}

\noindent
However, in the general case, the situation is more complicated. So, we need a notion of $\mathcal{S}$-gauge symmetries to gauge both $\mathcal{S}$-sections and solutions to spatial equations. 
\textit{From now on, we consider $\mathcal{S}$-stationary points of Lagrangian equations up to $(\boldsymbol \ell, \mathcal{S})$-gauge equivalence}. 

\section{Examples}{\label{Exam}}

We now consider several examples demonstrating how the Lagrangian formalism takes the form of the Hamiltonian formalism at the level of the intrinsic geometry of differential equations.

\subsection{Wave equation}

Let us consider the wave equation
\begin{align*}
u_{xy} = 0\,.
\end{align*}
Suppose that $\mathcal{S}$ is the lift of the characteristic distribution $\ker dy$. One can treat $x$, $y$, $u$, $u_x$, $u_y$, $u_{xx}$, $u_{yy}$, $u_{xxx}$, $u_{yyy}$, $\ldots$ as coordinates on the infinite prolongation $\mathcal{E}$ of the wave equation. 
Any vector field of the form
\begin{align*}
Y_{\varphi} = \varphi_0\,\partial_u + \varphi_1\partial_{u_y} + \varphi_2\,\partial_{u_{yy}} + \varphi_3\,\partial_{u_{yyy}} + \ldots
\end{align*}
is an $(\boldsymbol \ell, \mathcal{S})$-gauge symmetry, where $\varphi_0, \varphi_1, \ldots$ are arbitrary smooth functions of a finite number of the arguments $y$, $u_y$, $u_{yy}$, $u_{yyy}$, \ldots, $\boldsymbol \ell$ is represented by the differential form $l\in \Lambda^2(\mathcal{E})$,
\begin{align*}
l = -\dfrac{u_x u_y}{2} dx\wedge dy - \dfrac{u_y}{2} \,\overline{\!\theta}_0\wedge dy - \dfrac{u_x}{2} dx\wedge \,\overline{\!\theta}_0\,,\qquad \,\overline{\!\theta}_0 = du - u_x dx - u_y dy\,.
\end{align*}
Indeed, the form $dl$ represents the same $\mathcal{S}$-presymplectic structure as
\begin{align*}
\omega = \dfrac{1}{2} \,\overline{\!\theta}_x\wedge \,\overline{\!\theta}_0\wedge dx\,.
\end{align*}
Here $\,\overline{\!\theta}_x = du_x - u_{xx} dx$. It is easy to see that $Y_{\varphi}\, \lrcorner\, \omega = \dfrac{\varphi_0}{2}\, dx\wedge \,\overline{\!\theta}_x\in d\Big(\dfrac{\varphi_0}{2}\ \overline{\!\theta}_0\Big) + \mathcal{S}^2\Lambda^2(\mathcal{E})$.

\vspace{1.5ex}

\remarka{If $\varphi_1 = \,\overline{\!D}_y(\varphi_0)$, $\varphi_2 = \,\overline{\!D}_y^{\,2}(\varphi_0)$, \ldots (here $\,\overline{\!D}_y = D_y|_{\mathcal{E}}$), then $Y_{\varphi}$ is a symmetry of the wave equation.
}

\vspace{1.5ex}

As shown in~\cite{Druzhkov3}, an $\mathcal{S}$-section $\sigma$
\begin{align*}
\sigma\colon\qquad u = f(x, y)\,,\qquad u_x = \partial_x f\,,\qquad u_y = h_1(y)\,,\qquad u_{xx} = \partial_x^2 f\,,\qquad u_{yy} = h_2(y)\,,\qquad \ldots
\end{align*}
is an $\mathcal{S}$-stationary point of $\boldsymbol \ell$ if and only if $\partial_x \partial_y f = 0$. Therefore, any $\mathcal{S}$-stationary point $\sigma$ can be transformed into a solution of the wave equation using the transformation $\Phi^1$, where $\Phi^{\mathcal{T}}$ denotes the flow of the $(\boldsymbol \ell, \mathcal{S})$-gauge symmetry $Y_{\varphi}$ for
\begin{align*}
\varphi_0 = 0\,,\qquad \varphi_1 = -h_1 + \partial_y f\,,\qquad \varphi_2 = -h_2 + \partial_y^2 f\,,\qquad \varphi_3 = -h_3 + \partial_y^3 f\,,\qquad \ldots
\end{align*}

\remarka{Functions of the form $h_i(y)$ arise from a general solution to the spatial equation. This spatial equation is given by the \textit{infinite} number of independent constraints, including $y_x = 0$.}

\vspace{1.5ex}

Thus, any $\mathcal{S}$-stationary point of the internal Lagrangian $\boldsymbol \ell$ is $(\boldsymbol \ell, \mathcal{S})$-gauge equivalent to a solution of the wave equation. Let us note that, in the case under consideration, the $(\boldsymbol \ell, \mathcal{S})$-gauge symmetries do not allow compactly supported perturbations of solutions to the spatial equation. Nonetheless, this example shows that, in a sense, some characteristic distributions have a gauge-like nature.

\subsection{Maxwell's equations}

Let us return to the consideration of Example~\ref{MaxExample}.
Any $\mathcal{S}$-section $\sigma$ has the form
\begin{align}
\sigma\colon \qquad
\begin{aligned}
&A^{\nu} = f^{\nu}\,,\qquad F^{0i} = g^i\,,\qquad \partial_0 A^0 = h^1\,,\qquad \partial_0^2 A^0 = h^2 \,,\qquad \ldots\\
&\partial_i A^{\nu} = \partial_i f^{\nu}\,,\qquad \ldots
\label{Maxsigma}
\end{aligned}
\end{align}
The functions $f^{\nu}, h^1, h^2, \ldots \in C^{\infty}(\mathbb{R}^n)$ can be chosen arbitrarily, while $g^i \in C^{\infty}(\mathbb{R}^n)$ must satisfy one constraint: $\partial_i g^i = 0$. Here we use the notation $\partial_{\mu} f^{\nu}$, $\partial_{\mu} g^{i}$, $\ldots$ for the partial derivatives $\partial_{x^\mu} f^{\nu}$, $\partial_{x^\mu} g^{i}$, \ldots, while $\partial_{i} A^{\nu}$, $\ldots$ denote coordinates on the $\mathcal{E}$.

\vspace{1.5ex}

\remarka{If $n = 4$, then $(\boldsymbol \ell, \mathcal{S})$-gauge equivalence classes of solutions to the spatial equation can be identified with tuples $(t_0; E_0; B_0)$, where $E_0$ and $B_0$ are instantaneous electric and magnetic fields (at $t = t_0$), respectively.}

\vspace{1.5ex}

We find
\begin{align*}
\int \sigma^*(l) = \int \Big(\dfrac{1}{2}g^i g_i - \dfrac{1}{4}(\partial_i f_j - \partial_j f_i)(\partial^i f^j - \partial^j f^i) - g_i (\partial^0 f^i - \partial^i f^0)\Big)d^n x\,.
\end{align*}
One can resolve the constraint $\partial_i g^i = 0$ in the following way:
$g^i = \partial_j r^{ij}$, where $r^{ij}\in C^{\infty}(\mathbb{R}^n)$ are arbitrary functions such that $r^{ij} = -r^{ji}$; accordingly,
\begin{align}
\int \sigma^*(l) = \int \Big(\dfrac{1}{2}\partial_k r^{ik} \partial^{j} r_{ij} - \dfrac{1}{4}(\partial_i f_j - \partial_j f_i)(\partial^i f^j - \partial^j f^i) - \partial^j r_{ij}(\partial^0 f^i - \partial^i f^0)\Big)d^n x\,.
\label{MaxNonl}
\end{align}

\remarka{The locality of the general solution to the constraint is convenient for working with boundary conditions. More specifically, for any compact oriented $n$-dimensional submanifold $N\subset \mathbb{R}^n$, we can take as variations $\delta f^{\nu}, \delta r^{ij}, \delta h^1, \delta h^2, \ldots\in C^{\infty}(\mathbb{R}^n)$ arbitrary functions that vanish with all their derivatives on $\partial N$ and such that $\delta r^{ij} = - \delta r^{ji}$.}

\vspace{1.5ex}
\noindent
The variational problem for action~\eqref{MaxNonl} is reduced to the corresponding Euler-Lagrange equations
\begin{align*}
\begin{aligned}
&\partial_0 \partial_j r^{ij} = \partial_j(\partial^i f^j - \partial^j f^i)\,,\\
&\partial_j\big(\partial^k r_{ik} - (\partial_0 f_i - \partial_i f_0)\big) = \partial_i\big(\partial^k r_{jk} - (\partial_0 f_j - \partial_j f_0)\big)\,.
\end{aligned}
\end{align*}
The latter equation is equivalent to the existence of a function $\lambda\in C^{\infty}(\mathbb{R}^n)$ such that
\begin{align*}
&\partial^k r_{ik} - (\partial_0 f_i - \partial_i f_0) = \partial_i \lambda\,.
\end{align*}
Thus, an $\mathcal{S}$-section $\sigma$ written in the form~\eqref{Maxsigma} is an $\mathcal{S}$-stationary point of the internal Lagrangian $\boldsymbol \ell$ if and only if there is a function $\lambda\in C^{\infty}(\mathbb{R}^n)$ such that $\sigma$ satisfies the equations
\begin{align*}
&\partial_0 g^i = \partial_j(\partial^i f^j - \partial^j f^i)\,,\\
&g^i = \partial^0 f^i - \partial^i (f^0 - \lambda)\,.
\end{align*}
Let us recall that the condition $\partial_i g^i = 0$ is satisfied for all $\mathcal{S}$-sections. Therefore, an $\mathcal{S}$-stationary point written in the form~\eqref{Maxsigma} can be transformed into a solution of Maxwell's equations using the transformation $\Phi^1$, where $\Phi^{\mathcal{T}}$ denotes the flow of the $(\boldsymbol \ell, \mathcal{S})$-gauge symmetry $X_{(0, 0, \varphi)}$\,,
\begin{align*}
\varphi^0 = -\lambda\,,\qquad \varphi^1 = -h^1 + \partial_0 (f^0 - \lambda)\,,\qquad \varphi^2 = -h^2 + \partial_0^2 (f^0 - \lambda)\,,\qquad \ldots
\end{align*}

Thereby, we can draw the following remarkable conclusion about the Maxwell equations and the spatial distribution under consideration. All $\mathcal{S}$-stationary points of the Maxwell system are $(\boldsymbol \ell, \mathcal{S})$-gauge equivalent to its solutions!

\vspace{1.5ex}

\remarka{Since Maxwell's equations are Lorentz-invariant, the same conclusion can be made for all spatial distributions that one can obtain from the $\mathcal{S}$ using Lorentz transformations.}

\subsection{Potential KdV}

Let us consider an example of a variational equation that is not a Lagrangian one. 
The potential KdV equation
\begin{align*}
u_t = 3u_x^2 + u_{xxx}
\end{align*}
admits the differential consequence $\mathrm{E}(L) = 0$, where
$$
L = \Big(\dfrac{u_x u_t}{2} - u_x^3 + \dfrac{u_{xx}^2}{2}\Big)dt\wedge dx\,.
$$
One can treat $t, x, u, u_x, u_{xx}, u_{xxx}, \ldots$ as coordinates on the infinite prolongation $\mathcal{E}$ of the potential KdV equation. The corresponding internal Lagrangian $\boldsymbol \ell$ is represented by the differential form
\begin{align*}
l = \Big(\dfrac{u_x (3u_x^2 + u_{xxx})}{2} - u_x^3 + \dfrac{u_{xx}^2}{2}\Big)dt\wedge dx - \dfrac{1}{2}(3u_x^2 + u_{xxx})\,dt\wedge \,\overline{\!\theta}_0 + u_{xx}\,dt\wedge \,\overline{\!\theta}_x + \dfrac{1}{2}u_x\, \,\overline{\!\theta}_0\wedge dx\,,
\end{align*}
where $\,\overline{\!\theta}_0 = du - u_x\, dx - (3u_x^2 + u_{xxx})\, dt$ and $\,\overline{\!\theta}_x = du_x - u_{xx}\, dx - (6u_xu_{xx} + u_{xxxx})\, dt$.

Let $\mathcal{S}$ be the lift of the characteristic distribution $\ker dt$.
The $\mathcal{S}$-presymplectic structure is produced by the differential form
\begin{align*}
\omega = \dfrac{1}{2}\,\overline{\!\theta}_x\wedge \,\overline{\!\theta}_0\wedge dx\,.
\end{align*}
Any $\mathcal{S}$-symmetry of the potential KdV equation has the form
$$
X = \varphi\, \partial_u + \,\overline{\! D}_x(\varphi)\partial_{u_x} + \,\overline{\! D}_x^{\,2}(\varphi)\partial_{u_{xx}} + \ldots\,,
$$
where $\varphi$ is a function on $\mathcal{E}$, $\,\overline{\! D}_x = D_x|_{\mathcal{E}}$. Then
\begin{align*}
X \lrcorner \, \omega = \dfrac{1}{2}\Big(\,\overline{\! D}_x(\varphi) \, \overline{\!\theta}_0 - \varphi \ \overline{\!\theta}_x  \Big)\wedge dx \in 
\,\overline{\! D}_x(\varphi)\,\overline{\!\theta}_0\wedge dx + d\Big(\dfrac{\varphi}{2} \ \overline{\!\theta}_0\Big) + \mathcal{S}^2\Lambda^2(\mathcal{E})
\end{align*}
and $(\boldsymbol \ell, \mathcal{S})$-gauge symmetries are given by functions of the form $\varphi = \varphi(t)$.

Any $\mathcal{S}$-section $\sigma$ has the form
$$
\sigma\colon \qquad u = f,\qquad u_x = \partial_x f,\qquad u_{xx} = \partial_x^2 f,\qquad u_{xxx} = \partial_x^3 f,\qquad \ldots\,,
$$
where $f\in C^{\infty}(\mathbb{R}^2)$ can be chosen arbitrarily. Since the pullback reads
\begin{align*}
\sigma^*(l) = \Big(\dfrac{\partial_x f \partial_t f}{2} - (\partial_x f)^3 + \dfrac{(\partial^{\kern 0.05em 2}_{x}f)^2}{2} \Big)dt\wedge dx\,,
\end{align*}
the corresponding $\mathcal{S}$-stationary points are described by the Euler-Lagrange equation
\begin{align*}
\partial_x \Big(\partial_t f - 3(\partial_x f)^2 - \partial_x^{\kern 0.05em 3} f\Big) = 0\,,
\end{align*}
which is equivalent to the existence of a function $g(t)$ such that $\partial_t f = 3(\partial_x f)^2 + \partial_x^{\kern 0.05em 3} f + g(t)$. Denote by $\Phi^{\mathcal{T}}_g$ the flow of the $(\boldsymbol \ell, \mathcal{S})$-gauge symmetry for $\varphi(t) = -\int_0^t g(\tau) d\tau$. Then the transformations $\Phi^1_g$ relate the corresponding $\mathcal{S}$-stationary points of $\boldsymbol \ell$ to solutions of the potential KdV equation.

\vspace{1.5ex}

\remarka{There is a differential covering~\cite{VinKr} that relates $\mathcal{E}$ to the infinite prolongation $\mathcal{E}'$ of the KdV equation $v_t = 6vv_x + v_{xxx}$\,. One can choose $t, x, v, v_x\,, v_{xx}\,, \ldots$ as coordinates on $\mathcal{E}'$ and consider the differential covering $\rho\colon \mathcal{E}\to \mathcal{E}'$ that is given by the formulae $v = u_x$\,, $v_x = u_{xx}$\,, $\ldots$ Note that $\rho_*(\mathcal{S})$ is a well-defined spatial distribution on $\mathcal{E}'$. The $\rho$ establishes the one-to-one correspondence between $(\boldsymbol \ell, \mathcal{S})$-gauge equivalence classes of $\mathcal{S}$-sections of the potential KdV equation and $\rho_*(\mathcal{S})$-sections of the KdV equation.
Moreover, $(\boldsymbol \ell, \mathcal{S})$-gauge equivalence classes of $\mathcal{S}$-stationary points of $\boldsymbol \ell$ are in one-to-one correspondence with solutions of the KdV equation.
Thus, in this example, $(\boldsymbol \ell, \mathcal{S})$-gauge symmetries lead to the description of evolution determined by another equation.
An alternative approach is to consider the spatial equation a non-gauge one and not to take $(\boldsymbol \ell, \mathcal{S})$-gauge symmetries into account.}

\section{Discussion}{\label{Diss}}

In the previous sections, we formulated the non-covariant canonical variational principle for equations viewed as bundles. In fact, the role of a bundle structure was not of great importance. For example, one can define spatial distributions as arbitrary $(n-1)$-dimensional involutive distributions such that all their planes are subspaces of the respective ($n$-dimensional) Cartan planes, and so on. What is interesting in itself is that bundle structures are not required at all to define basic geometric structures on a differential equation.

Surprisingly, temporal parts of space+time decompositions do not participate in the canonical variational principle formulation.
Spatial equations play a more significant role. Namely, in addition to a simple physical interpretation, they allow us to introduce equivalence relations on reasonable classes of submanifolds in which internal Lagrangians can be varied invariantly. This is our motivation to consider them. However, they make the whole construction non-covariant. Essentially, they are the only additional structures involved in the construction. It turns out that it is possible to formulate a variational principle that does not rely on additional structures. So, to top it all, let us briefly formulate a covariant version of the canonical variational principle.

\subsection{Covariant canonical variational principle}

Let $\pi_{\mathcal{E}}\colon \mathcal{E}\to M^n$ be a differential equation. Suppose $\boldsymbol \ell$ is an internal Lagrangian of $\mathcal{E}$ represented by a differential form $l\in \Lambda^n(\mathcal{E})$.

\vspace{1.5ex}

\definitiona{A section $\sigma$ of the bundle $\pi_{\mathcal{E}}$ is an \textit{almost solution} (or an \textit{almost Cartan section}) if for each $x\in M$, 
\begin{align*}
\dim \big(d\sigma_x (T_{x} M) \cap \mathcal{C}_{\sigma(x)}\big) \geqslant n-1.
\end{align*}
}

\definitiona{A mapping $\gamma\colon \mathbb{R}\times M\to \mathcal{E}$ is a \textit{path in almost solutions} of $\pi_{\mathcal{E}}$ if the mappings
\begin{align*}
\gamma(\tau)\colon x\mapsto \gamma(\tau, x)
\end{align*}
are almost solutions of $\pi_{\mathcal{E}}$ for all $\tau\in\mathbb{R}$.
}

\vspace{1.5ex}

Let us assume that each spatial distribution on $\mathcal{E}$ is associated with a group of its spatial-gauge transformations.
We can introduce the following equivalence relation on the set of almost solutions.

\vspace{1.5ex}

\definitiona{Almost solutions $\sigma$ and $\sigma'$ of $\pi_{\mathcal{E}}$ are \textit{almost gauge equivalent} if there exist diffeomorphisms $f_1, \ldots, f_k\colon \mathcal{E}\to \mathcal{E}$ such that\\
\textbf{1)} each $f_i$ is an $\mathcal{S}_i$-gauge transformation, where $\mathcal{S}_i$ is a spatial distribution;\quad \textbf{2)} $\sigma$ is an $\mathcal{S}_1$-section;\\
\textbf{3)} $f_i\circ\ldots\circ f_1\circ \sigma$ is an $\mathcal{S}_{i+1}$-section for $i = 1, \ldots, k-1$;\quad \textbf{4)} $\sigma' = f_k\circ \ldots \circ f_2 \circ f_1 \circ \sigma$.}

\vspace{1.5ex}

\noindent
An almost solution $\sigma'$ can be an $\mathcal{S}$-section for several spatial distributions taken as $\mathcal{S}$ (it can define a local solution on an open subset of $M$). Since we consider all spatial distributions on an equal basis, such compositions $f_k\circ \ldots \circ f_2 \circ f_1 \circ \sigma$ are necessary to get an equivalence relation (because $\sigma \sim \sigma'$ and $\sigma' \sim \sigma''$ implies $\sigma \sim \sigma''$).

\vspace{1.5ex}

\definitiona{An almost solution $\sigma$ is a \textit{stationary point} of $\boldsymbol \ell$ if for any compact oriented $n$-dimensional submanifold $N\subset M$, the relation
\begin{align*}
\dfrac{d}{d\tau}\Big|_{\tau = 0}\int_N \gamma(\tau)^*(l) = 0
\end{align*}
holds for each path $\gamma$ in almost solutions such that $\gamma(0) = \sigma$ and all points of the boundary $\partial N$ are fixed (i.e., for each $x\in\partial N$, the condition $\gamma(\tau, x) = \gamma(0, x)$ is satisfied for all $\tau\in\mathbb{R}$).
}

\vspace{1.5ex}

\noindent
We say that an almost gauge equivalence class of almost solutions \textit{satisfies the covariant canonical variational principle} if it can be represented by a stationary point of $\boldsymbol \ell$. Again, the choice of a representative of $\boldsymbol \ell$ has no impact~\cite{Druzhkov3}. All solutions of a variational equation produce almost gauge equivalence classes that satisfy the covariant canonical variational principle.

\vspace{1.5ex}

\remarka{If $\boldsymbol L$ is an element of group~\eqref{IntSepLag} and $M$ is compact and oriented, then the action
\begin{align*}
\sigma\mapsto \int_M \sigma^*(\boldsymbol L)
\end{align*}
is well-defined on almost solutions such that $d\sigma_x(T_x\,\partial M)\subset \mathcal{C}_{\sigma(x)}$ for each $x\in \partial M$.
}

\vspace{3.0ex}

\centerline{\bf{\Large Acknowledgments}}

\vspace{2.0ex}

The author is grateful to M.~Grigoriev for significant discussions. The author thanks University of Saskatchewan for hospitality, and Prof. Alexey Shevyakov for financial support through the NSERC grant RGPIN 04308-2024.

\end{document}